# Measurement of the parity-violating triton emission asymmetry in the reaction $^6$Li(n,α)$^3$H with polarised cold neutrons

V. A. Vesna[1], Yu. M. Gledenov[2], V. V. Nesvizhevsky[3], A. K. Petoukhov[3], P. V. Sedyshev[2], T. Soldner[3], O. Zimmer[3,4], E. V. Shulgina[1]

[1] Petersburg Nuclear Physics Institute RAS, 188300 Gatchina, Leningrad Reg., Russia
[2] Joint Institute for Nuclear Research, 141980 Dubna, Moscow Reg., Russia
[3] Institut Laue-Langevin, F-38042 Grenoble Cedex 9, France
[4] Physik-Department E18, TU München, D-85748 Garching, Germany

**Abstract**

We describe measurements of the parity-violating (P-odd) triton emission asymmetry coefficient $a_{\text{P-odd}}$ in the $^6$Li(n,α)$^3$H reaction with polarised cold neutrons. Experiments were carried out at the Petersburg Nuclear Physics Institute (Gatchina, Russia) and at the Institut Laue-Langevin (Grenoble, France). We employed an ionisation chamber in a configuration allowing us to suppress the left-right asymmetry well below $10^{-8}$. A test for a false asymmetry due to eventual target impurities ("zero test") resulted in $a_{\text{0-test}} = (0.0 \pm 0.5) \times 10^{-8}$. As final result we obtained $a_{\text{P-odd}} = (-8.6 \pm 2.0) \times 10^{-8}$.

**Introduction**

Neutral weak currents were predicted by the unified theory of electromagnetic and weak interactions and discovered in 1973 in scattering of muon-neutrinos by electrons or nuclei [1, 2, 3]. In 1978 a neutral weak current was observed in low-energy physics as a P-odd rotation of the polarisation of a laser beam passing through atomic bismuth vapours [4]. The existence of neutral currents in lepton-lepton and lepton-nucleon interactions led to numerous attempts to discover experimentally neutral currents also in nucleon-nucleon (NN) interactions in nuclear reactions. As the weak interaction is too short-ranged for a direct exchange of weak bosons between the nucleons (hard core repulsion due to the strong interaction), NN processes are dominated by strong interaction. To describe the contribution of weak interaction, the problem is often parameterised in terms of single (π, ρ, ω) and multiple (ππ) meson exchange with one weak and one or more strong vertices (see [5, 6] for reviews). The physics of weak boson exchange (W$^\pm$, Z$^0$) is thus hidden in the weak vertex. Within this parameterisation, the neutral and the charged weak currents are described by the coupling constants $f_\pi$ and $h_\rho^0$, respectively. Several models attempt to link these couplings to more fundamental theory. The quark model by Desplanques, Donoughe and Holstein (DDH) predicts as "best values" $f_\pi = 4.6 \times 10^{-7}$ and $h_\rho^0 = -11.4 \times 10^{-7}$ within "allowed ranges" [7]. The soliton model of the nucleon by Kaiser and Meissner (KM) [8, 9] predicts lower values for these couplings. This reflects the difficulties of these models.

In order to get experimental information on the weak contributions to NN processes, the violation of spatial parity, specific to weak interaction, is commonly used as filter against the otherwise dominating strong interaction. Parity violation in NN interactions has been observed in various processes involving few-nucleon systems and more complex nuclei. While the observation of parity violation in proton-proton scattering is a manifestation of charged weak currents in accordance with theory [10, 11], no observation has yet been claimed for neutral weak current contributions in NN-interaction in nuclei.

A popular system for the theoretical and experimental investigation of P-odd effects is the radiative neutron capture by protons, n(p,γ)d. As shown in ref. [6], the coefficient $A_\gamma$ of the P-odd asymmetry for γ emission depends on the weak neutral current alone, $A_\gamma = -0.11 f_\pi$. The best value of the DDH model for $f_\pi$ corresponds to a P-odd asymmetry in the order of $2\times10^{-8}$. The model of KM predicts an even smaller value. To perform a statistically significant measurement of such a tiny observable, the asymmetry needs to be measured with a sensitivity of at least $5\times10^{-9}$. Such sensitivity has never been achieved experimentally despite intense efforts [12, 13]. A significant experimental difficulty is the cross section for neutron absorption being much smaller than for scattering.

Much larger parity violating effects were observed in neutron-induced reactions involving medium size and heavy nuclei, where peculiar enhancement effects may occur [14]. However, due to limited knowledge of nuclear wave functions the extraction of weak interaction constants from these reactions seems impossible [6]. On the other hand, considerable progress has been made in the theoretical description of light nuclear systems involving up to 10 nucleons, starting from phenomenological potential models for NN interactions and including three-body forces. It is a hope that in foreseeable future effective field theoretical approaches will challenge experimentalists with a parameter-free description of light nuclear systems. Still lacking such a description, we analyse our data using a cluster model. In this approach the interaction of neutrons with light nuclei is considered as a few-nucleon reaction, influenced by the potential of one or a few α-particles. Then the problem can be formulated in terms of constants of the meson-nucleon interaction. In [15] this model has been employed to the following system of light nuclei for excitation energies below 20-25 MeV:

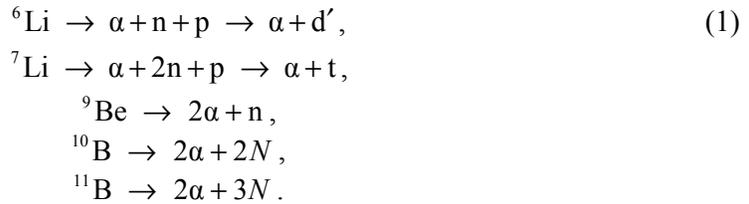

$$^{6}\text{Li} \to \alpha+n+p \to \alpha+d', \quad (1)$$
$$^{7}\text{Li} \to \alpha+2n+p \to \alpha+t,$$
$$^{9}\text{Be} \to 2\alpha+n,$$
$$^{10}\text{B} \to 2\alpha+2N,$$
$$^{11}\text{B} \to 2\alpha+3N.$$

As example, the model predicts 70% clusterization of the $^6$Li nucleus in the form of an α particle and a deuteron d′ deformed by the α particle potential. Therefore the reaction $^{6}\text{Li}(n,\alpha)^{3}\text{H}$ can be considered as a 3-body reaction between a neutron, an α particle and a deuteron d′:

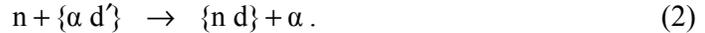

$$n + \{\alpha\, d'\} \to \{n\, d\} + \alpha. \quad (2)$$

The P-odd asymmetry coefficient $a_{\text{P-odd}}$ for triton emission in $^{6}\text{Li}(n,\alpha)^{3}\text{H}$ was calculated in [15] in terms of the constants $h_\rho^0$ and $f_\pi$:

$$a_{\text{P-odd}} \approx (-0.45 f_\pi + 0.06 h_\rho^0) = -2.8 \times 10^{-7}. \quad (3)$$

The numerical value employs the DDH "best values" for the weak constants. The P-odd asymmetry coefficient for emission of γ–quanta in the reaction $^{10}B(n,\alpha)^7Li^*$, $^7Li^* \to {}^7Li+\gamma$ [16] was calculated as:

$$a_{\text{P-odd}} \approx (-0.078 f_\pi + 0.028 h_\rho^0) = -0.7 \times 10^{-7}. \qquad (4)$$

Hence, a measurement of the asymmetry coefficients for these two reactions will allow us to derive values for the neutral and charged weak interaction constants. In our experimental programme we investigate the P-odd asymmetries in the reactions $^6Li(n,\alpha)^3H$ (asymmetry of triton emission) and $^{10}B(n,\alpha)^7Li^*$, $^7Li^* \to {}^7Li+\gamma$ ($E_\gamma$ = 0.48 MeV) (asymmetry of γ-[16] or α-emission [17]). Compared to n(p,γ)d these reactions have the advantage that the expected asymmetry values are significantly higher. The experiment is also much easier to implement, due to large neutron absorption cross sections in the order 1000 barns, which allows us to make effective use of the available neutron flux. Present intense neutron sources provide sufficient statistical sensitivity for the detection of non-zero P-odd effects even in light nuclear systems. In this paper we present the final results of the $^6Li(n,\alpha)^3H$ experiments.

## Principles of the experiment

The measurement of the P-odd asymmetry in $^6Li(n,\alpha)^3H$ with a precision of $10^{-8}$ faces two principal challenges: to obtain the necessary statistics and to suppress possible false effects.

In order to collect $10^{16}$ events, a strong neutron beam has to be used, providing count rates of typically $10^{10}$ s$^{-1}$. At such rates, direct counting is not possible anymore and an integral technique of the event detection was used. The integral method of the P-odd asymmetry measurements was proposed first by V.M. Lobashev [18, 19]. To distinguish α and triton, one can use the different range of these two particles in material and stop the α in a foil of adequate thickness. The triton is then detected in an ionisation chamber. In the current integrated method, the signal results from the energy deposited by the different tritons averaged over the accepted emission angles.

The left-right asymmetry coefficient $a_{LR}$ of the reaction, describing the triple correlation $\vec{\sigma}_n \cdot [\vec{p}_n \times \vec{p}_t]$ of neutron spin and the momentums of neutron and triton, is $a_{LR}$ = (1.06 ± 0.04)×10$^{-4}$ [20]. False effects due to an admixture of this asymmetry have to be suppressed by choosing the average emission angle of the triton, the neutron spin, and the neutron momentum all parallel with a precision of ε~5×10$^{-3}$. Then the false effect, proportional to ε$^2$, will not exceed 2.5×10$^{-9}$. Variations in the reactor power need to be compensated by measuring neutron spin and triton emission parallel and anti-parallel at the same time. False effects due to reactions with nuclei in the detector material or due to impurities of the target have to be avoided by proper choice of materials and were investigated by a zero measurement.

## The neutron beams

We performed three measurements of the P-odd asymmetry of triton emission in the reaction $^6$Li(n,α)$^3$H. A first run was carried out at the vertical neutron beam at the VVR-M reactor in the Petersburg Nuclear Physics Institute (PNPI, Gatchina, Russia) [21]. Two more runs were performed at the cold neutron beam facility PF1B at the Institut Laue-Langevin (ILL, Grenoble, France). There the mean neutron wavelength is 4.7 Å. Further beam characteristics can be found in ref. [22]. The first run at PF1B was carried out using a focusing super-mirror neutron polarizer with a cross section of 100×50 mm$^2$. In the second run a super-mirror polarizer with a cross-section of 80×80 mm$^2$ (beam width 60 mm) was used. The total flux of polarised neutrons was about (4–5)×10$^{10}$ s$^{-1}$. The neutron spin direction could be inverted by an adiabatic spin flipper [23]. The neutron beam polarisation after passage through the ionisation chamber was checked using a super mirror polarizer as analyser. The product of polarisation, analysing power and flipping efficiency was $P = 0.88$ in the first run at the ILL, and $P = 0.94$ in the second run.

The width and the height of the neutron beam were reduced inside the ionisation chamber to a size smaller than the targets using three collimators made of $^6$LiF. The vertical deviation of the neutron beam from the chamber axis was smaller than 5 mm/m. The chamber was installed on concrete blocks in such a way that the horizontal deviation of the beam from the ionisation chamber axis was also smaller than 5 mm/m.

### The ionization chamber

The total kinetic energy release of 4.78 MeV in the reaction $^6$Li(n,α)$^3$H is shared among the reaction products as $E_\alpha$ = 2.05 MeV and $E_t$ = 2.73 MeV. Since the penetration length for the tritons is 5 times larger than for the α particles, the latter can be stopped completely by an Al foil without stopping the tritons. No further discrimination is then needed which allows us to use the integral detection method in an ionisation chamber. A target consisted of a $^6$LiF layer with a thickness of (400–500) μg/cm$^2$, evaporated onto an aluminium foil with thickness 14 μm and covered by an aluminium foil of the same thickness. Tritons entering a foil under flat angles may be stopped. This effect was taking into account in the average cosine of the triton emission angle $\langle\cos(\vec{\sigma}_n,\vec{p}_t)\rangle = 0.75$, calculated by Monte Carlo simulations.

Figure 1 shows the ionisation chamber (see also [17]). For high statistical sensitivity the longitudinally polarised neutron beam was passing through a total of 24 targets mounted perpendicular to the beam, absorbing about 60 % of the neutron beam intensity. A detector on each side of each target created a local "double chamber". One side of each double chamber detected tritons emitted in the neutron beam direction ("forward"), the other one in the opposite, backward direction. Due to the different orientations of the triton momentums $\vec{p}_t$ with respect the neutron spin $\vec{\sigma}_n$ the asymmetries measured in forward and backward directions have opposite signs. The electrodes attached to all the "forward" sections of the detector's double chambers were connected to a common "forward" preamplifier; and the electrodes attached to the "backward" sections to a separate "backward" preamplifier. The ionisation chamber was filled with gaseous Ar (absolute pressure 2.4 bar in the first experiment, 1.9 bar in the second experiment at the ILL). The path of the tritons emitted from a target was shorter than the 21 mm thick sensitive zone in each detector.

A solenoid fixed to the ionisation chamber provided a magnetic field parallel to the axis of the neutron beam, with directional accuracy better than 5×10$^{-3}$, providing the necessary suppression of the left-right asymmetry.

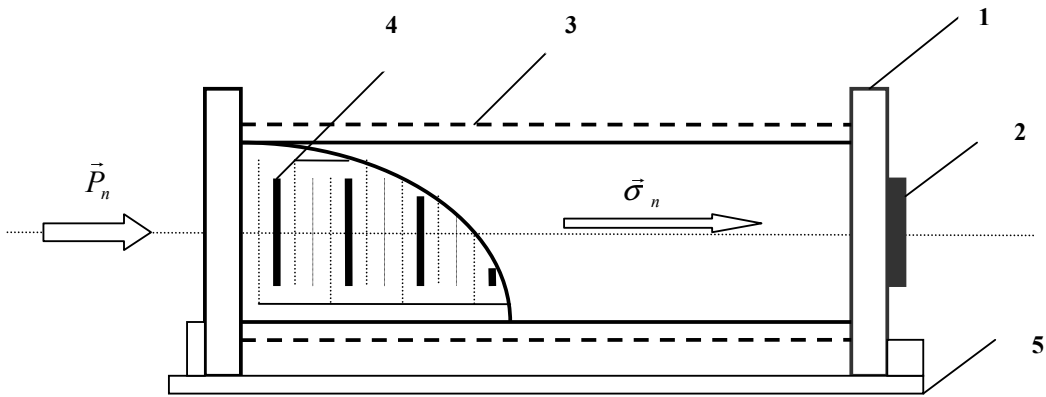

Fig. 1. Sketch of the ionisation chamber placement in the neutron beam:
1 – ionisation chamber; 2 – $^6$LiF beam-stop for absorption of neutrons at the chamber exit;
3 – solenoid for guiding the magnetic field; 4 – grids and $^6$LiF targets;
5 – base plate of the ionisation chamber.

## Electronics

The two preamplifiers converted the electric currents from the "forward" and the "backward" chambers to electrical voltages, which are proportional to the neutron flux. Figure 2 shows these voltages, as a function of the voltage at the high-voltage detector electrodes.

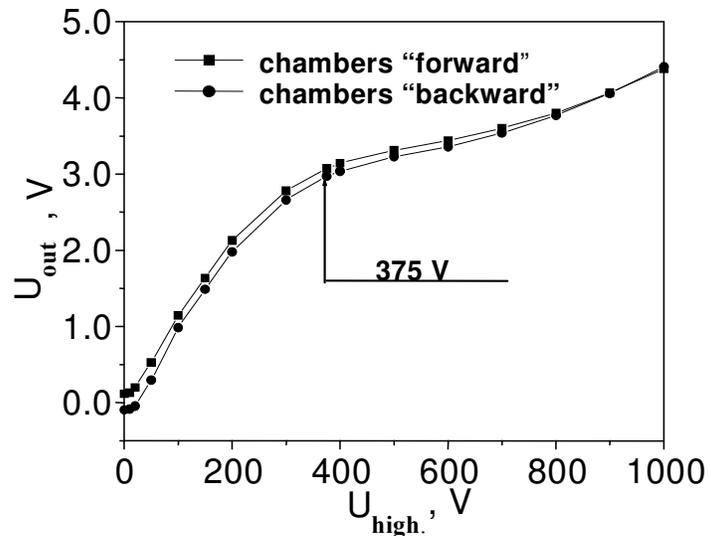

Fig. 2. Signals at the pre-amplifier output are shown as a function of the voltage at the high-voltage detector electrodes.

The preamplifiers decomposed the electrical signal into a constant and a variable component. The RC-circuit for the variable component was chosen in such a way, that it transmits a signal with the frequency equal to the flipper state switching frequency (~0.5 Hz) without any distortions of the signal. The variable component was amplified $K$~30 times and sent to an integrator [17]. The amplification coefficient was measured by applying a rectangular signal of known amplitude to the input. The constant component was measured for each series (once every 256 s) and saved in the computer memory. An operational amplifier [17] equipped with a capacitor was used as integrator.

A programmable analog-to-digital converter (a multi-function card for data acquisition: ACL-8112pg/ ADLink Technology Inc. with 12 digits) read the signals at the integrator output and sent them to the computer. The integration and other time intervals were set in the programmable counter-timers of an ACL-7120 card of ADLink Technology Inc. This card was also equipped with 32 TTL-compatible digital inputs/outputs used to control the experiment. Four reversible independent programmable counters of the card were used for setting the time intervals for the experiment (fig. 3). These counters are based on generators with a frequency of 50 kHz, allowing for a relative accuracy of $2\times10^{-5}$ for the time intervals. The electronics and the experiment control system are described in detail in refs. [17, 24].

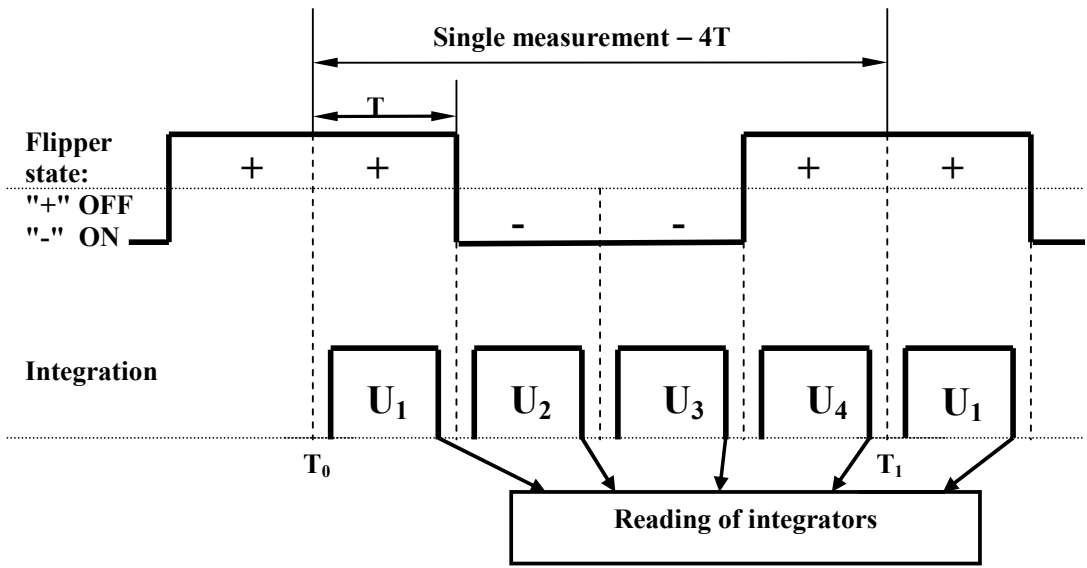

Fig. 3. The time intervals in the experiment.

## The measuring procedure and data treatment

To achieve an accuracy of ~$10^{-8}$ in the asymmetry measurement any fluctuations in the electronics or reactor neutron flux have to be minimised, as well as any interference from external electric signals or other false effects.

To compensate for fluctuations in reactor power we used special measuring procedures involving a pair of detectors for every target. Both detection chambers ("forward" and

"backward") simultaneously measure the same process but give opposite signs for the asymmetry effect (see fig. 1); fluctuations in reactor power will have an identical impact in both chambers and will carry the same sign.

In the integral method electrical signals can be presented as the sum of their constant and variable components. The "number of events" in a time interval is proportional to the sum of the variable $U/K$ and constant $U_C$ components of the signal, integrated over this interval; therefore the asymmetry coefficient $a_\text{P-odd}$ is:

$$a_\text{P-odd} = \frac{(U_C^+ + U^+/K) - (U_C^- + U^-/K)}{(U_C^+ + U^+/K) + (U_C^- + U^-/K)}, \quad (5)$$

where $U_C^+, U_C^-, U^+, U^-$ are the constant and variable components of the signal for two opposite neutron polarisations, relative to the detected particles' momentum in the respective chamber. The coefficient $K$ describes the amplification of the variable component of the signal. As $U_C \gg U$ and $U_C^+ \cong U_C^- = U_C$, the normalised asymmetry coefficient is given by:

$$a_\text{P-odd} = (U^+ - U^-)/(2KU_C).$$

For every detection channel, four consecutive voltages at the integrators $U_1^+, U_2^-, U_3^-, U_4^+$ (fig. 3) are combined in a "single measurement" and added as follows: $U^+ = U_1^+ + U_4^+$, $U^- = U_2^- + U_3^-$. These combinations allow us to suppress linear drifts. The asymmetry was calculated for each single measurement in every detector. In both channels the results of $N$ subsequent measurements in a series were summed; the constant component of a signal was measured once per series.

The effects for single channels 1 and 2 were calculated using the following formulas:

$$\left.\begin{array}{l} \overline{\alpha}^{(1),(2)} = \dfrac{1}{N}\sum_{i=1}^{N}\alpha_i^{(1),(2)} \\[4pt] \alpha_i^{(1),(2)} = \left(U_i^{+,(1),(2)} - U_i^{-,(1),(2)}\right)/2 \\[4pt] U_i^{+,(1),(2)} = U_{i,1}^{+,(1),(2)} + U_{i,4}^{+,(1),(2)}, \quad U_i^{-,(1),(2)} = U_{i,2}^{-,(1),(2)} + U_{i,3}^{-,(1),(2)}, \quad i = 1 \div N \\[4pt] D(\overline{\alpha}^{(1),(2)}) = \dfrac{1}{N(N-1)}\sum_{i=1}^{N}\left(\alpha_i^{(1),(2)} - \overline{\alpha}^{(1),(2)}\right)^2, \quad \sigma(\overline{\alpha}^{(1),(2)}) = \sqrt{D(\overline{\alpha}^{(1),(2)})} \end{array}\right\}, \quad (6)$$

$$\left.\begin{array}{l} \overline{\delta}^{(1),(2)} = \dfrac{\overline{\alpha}^{(1),(2)}}{K_{(1),(2)}U_{C,(1),(2)}} \\[8pt] \sigma(\overline{\delta}^{(1),(2)}) = \dfrac{\sigma(\overline{\alpha}^{(1),(2)})}{K_{(1),(2)}U_{C,(1),(2)}} \end{array}\right\}. \quad (7)$$

where $\overline{\alpha}^{(1),(2)}$, $\overline{\delta}^{(1),(2)}$ are the absolute (t.i. not normalised) and relative (normalised) effects for channels 1 and 2; $U_{i,1}^{+,(1)}, U_{i,2}^{-,(1)}, U_{i,3}^{-,(1)}, U_{i,4}^{+,(1)}$ are the measured voltages at the 1st integrator output ; $U_{i,1}^{+,(2)}, U_{i,2}^{-,(2)}, U_{i,3}^{-,(2)}, U_{i,4}^{+,(2)}$ are the measured voltages at the 2nd integrator output; $\sigma$ is the uncertainty in the average values of the effects; $K_{(1)}$, $K_{(2)}$ denote the amplification factors for the 1st and 2nd channel.

Taking advantage of the double detector chambers, one can calculate values compensated for fluctuations, e.g., of reactor power.

$$\left.\begin{array}{l}\bar{\alpha}_{\text{comp}} = \dfrac{1}{N}\left(\sum\limits_{i=1}^{N}\alpha_i^{(1)} - L\sum\limits_{i=1}^{N}\alpha_i^{(2)}\right) \\ D(\bar{\alpha}_{\text{comp}}) = \dfrac{1}{N(N-1)}\sum\limits_{i=1}^{N}\left((\alpha_i^{(1)} - L\alpha_i^{(2)}) - \bar{\alpha}_{\text{comp}}\right)^2 \\ \sigma(\bar{\alpha}_{\text{comp}}) = \sqrt{D(\bar{\alpha}_{\text{comp}})}\end{array}\right\}, \quad (8)$$

$$\left.\begin{array}{l}\bar{\delta}_{\text{comp}} = \dfrac{\bar{\alpha}_{\text{comp}}}{K\left(U_{C,(1)} + U_{C,(2)}\right)} \\ \sigma(\bar{\delta}_{\text{comp}}) = \dfrac{\sigma(\bar{\alpha}_{\text{comp}})}{K\left(U_{C,(1)} + U_{C,(2)}\right)} \\ K = \dfrac{K_{(1)} + LK_{(2)}}{2}\end{array}\right\}. \quad (9)$$

where $\bar{\alpha}_{\text{comp}}$, $\bar{\delta}_{\text{comp}}$ are the absolute and relative effects after compensation, $L$ is the compensation coefficient defined below; $U_{C,(1)}$ and $U_{C,(2)}$ are the constant components of the 1st and 2nd pre-amplifier signals.

The calculated effects $\alpha_i^{(1)}$ and $\alpha_i^{(2)}$ for the two detectors are of opposite sign; these values are measured synchronously. Therefore by calculating $\bar{\alpha}_{\text{comp}}$ the asymmetry value is doubled, and the effect of fluctuations in the reactor power is subtracted due to subtraction of one value from the other.

The compensation coefficient $L$ is calculated for every series of single measurements with the condition that the variation $D(\bar{\alpha}_{\text{comp}})$ of the average value of the absolute effect $\bar{\alpha}_{\text{comp}}$ (8) is minimal; it is defined as follows [17, 22]:

$$L = \dfrac{\sum\limits_{i=1\div N}\alpha_i^{(1)}\alpha_i^{(2)} - \dfrac{1}{N}\sum\limits_{i=1\div N}\alpha_i^{(1)}\sum\limits_{i=1\div N}\alpha_i^{(2)}}{\sum\limits_{i=1\div N}(\alpha_i^{(2)})^2 - \dfrac{1}{N}\left(\sum\limits_{i=1\div N}\alpha_i^{(2)}\right)^2}. \quad (10)$$

The integration interval was equal to 0.8 sec, and each series contained 64 single measurements (see Fig. 3). The variations of the compensated values was smaller by a factor of 40 – 100 than the variations of the non-compensated values.

The final result is an averaged value over a large number $n$ of results for series $\alpha_i \pm \sigma_i$:

$$\bar{A} = \dfrac{\sum\limits_{i=1}^{n}p_i\alpha_i}{\sum\limits_{i=1}^{n}p_i}, \quad p_i = \dfrac{1}{\sigma_i^2}. \quad (11)$$

with uncertainty $\qquad \sigma(\bar{A}) = \dfrac{1}{\sqrt{\sum\limits_{i=1}^{n}p_i}}.$

To further compensate false asymmetries we changed the direction of the guiding magnetic field and measured an equal number of series for both directions. For the averaged values we took into account the directions reversal of the real P-odd effect due to the reversal of the guiding magnetic field. In the first experiment [25] the field direction was reversed every 30 minutes, in the last experiment [26] every 4 minutes.

We emphasise that the three stages of signal evaluation all work with differences:
1. between variable parts of the signals (absolute effect of the P-odd asymmetry) $\alpha = U^+ - U^-$ for opposite neutron spin polarisations; these differences are calculated for every pair of measurements for both detectors.
2. between the absolute effects for the two detectors $\alpha_i = \alpha_i^{(1)} - L \cdot \alpha_i^{(2)}$, $i = 1 \div N$. Since the P-odd asymmetries in the detectors have opposite signs, the effect is doubled. These differences are calculated for each direction of the guiding magnetic field (referred to as "→" and "←").
3. between the effects for each direction of the guiding magnetic field: $\alpha_i(\rightarrow) - \alpha_i(\leftarrow)$, $i = 1 \div N$. P-odd effects are added in this case, because they have opposite signs.

At each of the 3 stages of calculation described above we subtract values measured for two opposite conditions. Effects of equal sign and equal size thus cancel, and the asymmetry persists. Taking the third difference, for instance, any influence of the guiding magnetic field on the currents in the chamber, or changes in neutron absorption as a function of the field direction would cancel unless the effect is due to P-odd effects from impurity nuclei.

This conclusion is valid as well for electromagnetically-induced false effects. Such influences were checked many times using electronic test signals as described in refs. [21, 25, 26]. For instance, a small effect was caused by switching the radio-frequency of the adiabatic spin-flipper on and off. After normalisation with the constant signal components and the coefficient of amplification the measured asymmetry contained a false shift of $-(0.4\pm1.5)\times10^{-9}$ [27]. This is considerably smaller than the experimental uncertainty. The identical treatment of the results for two detectors and for the two directions of guiding magnetic field thus allows us to avoid completely any influence of parasitic electromagnetic effects.

The Earth's magnetic field and stationary magnetic fields from other experimental installations are not shielded. These magnetic fields could increase a contribution to the P-odd asymmetry of the left-right asymmetry, which has the same sign for opposite directions of the guiding magnetic field [21]. If the number of series for the opposite field directions is equal and the results of these measurements are subtracted from each other, as described above, the contribution of this effect is suppressed.

**Measurement of a contribution of the left-right asymmetry and of the P-odd effect**

As mentioned earlier, the left-right asymmetry for the correlation $\vec{\sigma}_n \cdot [\vec{p}_n \times \vec{p}_t]$ is $a_{LR} = (1.06 \pm 0.04)\times10^{-4}$ [20], and a contribution to the measurement of the P-odd asymmetry has to be avoided. Therefore we performed several test experiments where we introduced intentionally a misalignment between $\vec{\sigma}_n$, $\vec{p}_n$ and $\vec{p}_t$.

By aligning the magnetic field perpendicular to the neutron beam we increased the sensitivity to a possible misalignment between $\vec{p}_n$ and $\vec{p}_t$. In this case, with the estimated precision of the alignment of $\vec{p}_n$ and $\vec{p}_t$, $\varepsilon \sim 5\times10^{-3}$, we expect a contribution of the left-right asymmetry of $\sim 5\times10^{-7}$ which can be easily measured. In a first series of measurements, we applied a vertical magnetic field and varied the angle of the chamber (i.e. the angle between $\vec{p}_n$ and $\vec{p}_t$) in the horizontal plane. Figure 4 shows the result [26]. The angle corresponding to zero left-right asymmetry contribution and the angle corresponding to the maximum neutron intensity (the maximum constant component) deviate by $10^{-2}$.

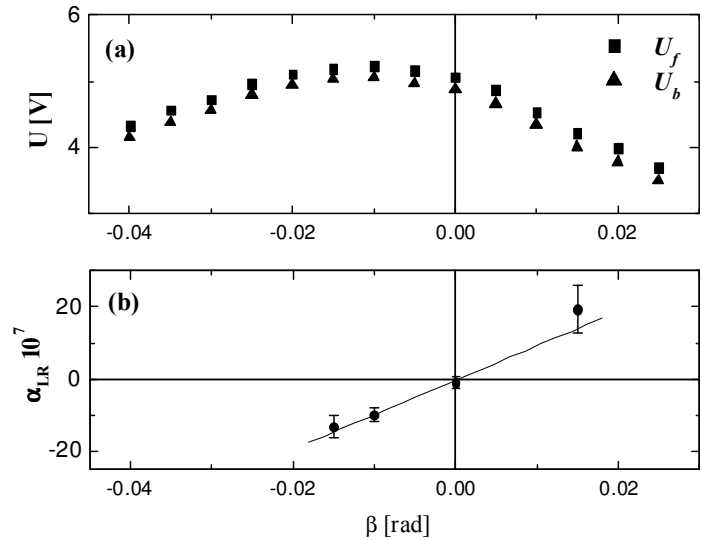

Fig. 4. (a) The constant components $U_f$ ("forward"), $U_b$ ("backward") and (b) the value of the left-right asymmetry coefficient $a_{LR}$ are shown as a function of the angle $\beta$ between the neutron momentum and the triton momentum. These measurements were performed with a vertical guiding magnetic field.

In a second measurement, we aligned the guiding magnetic field horizontally and perpendicular to the neutron beam [25], in order to check for a contribution of left-right asymmetry because of a vertical inclination of the chamber. For the best alignment of the chamber relative to the beam in terms of intensity, we obtained the asymmetry $a_{LR} = (4.9 \pm 1.9)\cdot 10^{-7}$.

With the magnetic field parallel to the neutron beam axis, the contribution of the left-right asymmetry to the P-odd effect will decrease by 2 orders of magnitude, as the direction of the guiding magnetic field is set by a solenoid fixed to the chamber with an accuracy of $\varepsilon \sim 5\times 10^{-3}$. From the test measurements we thus conclude that the influence of the left-right asymmetry to the measurement of the P-odd effect will be well below $10^{-8}$.

For an additional experimental verification of the contribution of the left-right asymmetry to the P-odd asymmetry in the latest experiment [26], the asymmetry was measured for three small angles between the neutron momentum and the chamber axis, but now with the longitudinal magnetic field used for the measurement of the P-odd effect. According to our calculations, the contribution of the left-right asymmetry to the P-odd effect should be suppressed by the precision of the alignment of the magnetic field, $5\times 10^{-3}$, compared to fig. 4(b), and not exceed $1\cdot 10^{-8}$ for the small angles used. Furthermore, such contribution would depend on the angle, as is clear from fig. 4(b). Table 1 shows the results for the three angles $\beta$ between the neutron momentum and the chamber axis. The angle $\beta=0$ corresponds to the minimum value of the left-right asymmetry (fig. 4). Obviously, the contribution of the left-right asymmetry to the measurement of the P-odd effect was smaller than the statistical uncertainties.

Table 1.

| β(rad) | $a_{\text{P-odd}}$ |
|---|---|
| ~ 0 | $-(10.8 \pm 4.4) \cdot 10^{-8}$ |
| - 0.01 | $-( 9.9 \pm 4.1) \cdot 10^{-8}$ |
| - 0.015 | $-( 7.4 \pm 4.2) \cdot 10^{-8}$ |

Average: $a_{\text{P-odd}} = -( 9.3 \pm 2.5) \cdot 10^{-8}$

Table 2 shows the results of the three measurements of the P-odd asymmetry coefficient in the reaction $^6\text{Li}(n,\alpha)^3\text{H}$ obtained by the authors. The results in tables 1 and 2 are corrected for the neutron polarisation $P$ and for the effective solid angle of triton detection:

Table 2.

| | $P\langle \cos(\vec{\sigma}_n, \vec{p}_t) \rangle$ | $a_{\text{P-odd}}$ |
|---|---|---|
| PNPI, vertical channel: | 0.66 | $- (5.4 \pm 6.0) \cdot 10^{-8}$ [19] |
| ILL, PF1B beam: | 0.66 | $- (8.1 \pm 3.9) \cdot 10^{-8}$ [23] |
| ILL, PF1B beam: | 0.70 | $- (9.3 \pm 2.5) \cdot 10^{-8}$ [24] |

Average: $a_{\text{P-odd}} = - (8.6 \pm 2.0) \cdot 10^{-8}$

### P-odd effects caused by impurity nuclei

Parity violating effects due to contaminant nuclei in the target cannot be compensated. A source of such background on the signal is the reaction of neutrons with the construction materials of the experimental installation, from which charged particles, γ-quanta and electrons from β-decay of radioactive nuclei are emitted. These false effects cannot be reliably estimated and cannot be tested in measurements with unpolarised neutrons.

A potential source of false effects is the asymmetry in the β-decay of nuclei with small lifetimes; in this case a non-zero nuclear polarisation after capture of a polarised neutron can persist until β-decay. The most abundant (~10%) impurity in the target material is $^7$Li, which produces $^8$Li ($T_{1/2} = 0.84$ s) by neutron capture. In [28] the β asymmetry of a pure $^7$Li sample was measured as $a = -0.08 \pm 0.01$. In our present setup, with an average energy of ~ 13 MeV the β-particle passes through nearly the entire sensitive volume of the ionisation chambers. The absorption is low and the energy is released in forward and in backward chambers. The P-odd effect due to β-particles from $^7$Li in the targets was estimated to be below $3\times10^{-9}$. The small value is due to the small neutron cross section of $^7$Li compared to that of $^6$Li; due to small fraction of $^7$Li in the target; and due to the small absorption of β-particles in the chamber.

The ionisation chambers are assembled using materials containing fluorine F. The halftime of $^{20}$F is $T_{1/2} = 11.16$ s; and the end-point energy of the β-particles is 5.4 MeV. One cannot exclude a P-odd effect with β-particles either. It is hardly possible to estimate the contribution of this reaction to the P-odd effect, given our poor understanding of the mass of $^{20}$F and the flux of neutrons scattered through the construction materials of the chamber. It is also true that the degree of nuclear polarisation after capture is uncertain. Therefore this and similar effects were investigated by the zero experiment described below.

In the reaction $^{35}$Cl(n,p)$^{35}$S a P-odd proton emission asymmetry was found with a large coefficient of $a = -(1.5 \pm 0.3) \times 10^{-4}$ [29]. The proton path in aluminium is 1.5 mg/cm$^2$. Thus, these protons can not traverse the aluminium foils surrounding the targets (their thickness of 14 μm corresponds to 3.78 mg/cm$^2$). Therefore protons from possible Cl impurities in the targets can not contribute to the P-odd effect. Even if the Cl admixture in the aluminium foils were as high as unrealistic 1%, the calculated ratio of chamber current due to protons to current due to tritons from $^6$Li would be as low as ~ 5.7·10$^{-10}$.

Calculations showed that an additional P-odd asymmetry due to γ-quanta from all known cases is lower than 10$^{-10}$.

The P-odd asymmetry for β-decay of various nuclei with long half-life in the construction materials (for instance, β-decay of aluminium) is not significant either, because those nuclei decay within a few minutes, and the interaction of the nuclear magnetic moment with the atomic environment immediately destroys nuclear orientation produced by polarised neutron capture.

As it is difficult to provide a complete estimate of all contributions to the P-odd effect from impurity nuclei, we performed a control experiment.

### A control experiment

Studies of tiny asymmetries usually include a control measurement ("zero experiment") with a polarised neutron beam, in which the products of the reaction are not detected.

Total absorption of the tritons and α particles is provided by additional 20 μm thick aluminium foil covering of the targets. All results in this section are given normalised to the constant component of the detector signal in the main experiment and to the solid angle and neutron polarisation, in order to make them directly comparable with the asymmetry in the main experiment. The result obtained at PNPI in Gatchina was $a_{0-\text{test}} = (2.0 \pm 1.7) \times 10^{-8}$ [21]. The zero-test carried out at the PF1B neutron beam at the ILL was performed with a neutron polarisation of more than 94% and yielded [27]:

$$a_{0-\text{test}} = -(0.0 \pm 0.5) \times 10^{-8}. \qquad (12)$$

Note that this measurement also tests for influences of electronics and noise from the flipper. In addition, we also carried out an asymmetry measurement with the neutron beam switched off which is sensitive to these latter effects, only. The result is [26]:

$$a_{\text{noise}} = -(0.6 \pm 0.5) \times 10^{-8}. \qquad (13)$$

Hence, all these false effects are at least 10 times smaller than the observed asymmetry.

### Discussion of the results and estimation of the weak neutral current

We measured the parity-violating (P-odd) triton emission asymmetry coefficient $a_{\text{P-odd}}$ in the $^6$Li(n,α)$^3$H reaction with polarised cold neutrons. It is equal to:

$$a_{\text{P-odd}} = (-8.6 \pm 2.0) \times 10^{-8}$$

We analyse this result within the nuclear cluster model. Using the formula (1) for the relation of the P-odd effect with the effective weak couplings and supposing that the charged weak current constant is given by the DDH best value $h_\rho^0 = -11.4\times10^{-7}$, the neutral weak constant $f_\pi$ and its uncertainty would be given by:

$$f_\pi \approx (0.4\pm0.4)\times10^{-7}.$$

At 90% confidence level, we obtain the following constraint for the neutral weak constant:

$$-0.3\times10^{-7} \leq f_\pi \leq 1.1\times10^{-7}.$$

This constraint based on the use of a nuclear cluster model is currently one of the most precise estimates available.

**References:**

[1] F.J. Hasert, H. Faissner, W. Krenz et al., Phys. Lett. B **46** (1973) 121.
[2] F.J. Hasert, S. Kabe, W. Krenz et al., Phys. Lett. B **46** (1973) 138.
[3] D. Cline, A.K. Mann, C. Rubbia, Sci. Am. **231**(6) (1974) 108.
[4] L.M. Barkov, M.S. Zolotarev, I.B. Khriplovich, Uspekhi Fiz. Nauk **132**(3) (1980) 409 (in Russian).
[5] E.G. Adelberger and W.C. Haxton, Ann. Rev. Nucl. Part. Sci. **35** (1985) 501.
[6] B. Desplanques, Phys. Rep., **297** (1998) 1.
[7] B. Desplanques, J.F. Donoughe and B.R. Holstein, Ann. Phys. **124** (1980) 449.
[8] N. Kaiser and U.G. Meissner, Nucl. Phys. A **489** (1988) 671.
[9] N. Kaiser and U.G. Meissner, Nucl. Phys. A **499** (1989) 699.
[10] I. Balzer, R. Hennek, Ch. Jachquemart et al., Phys. Rev., **C30** (1984) 1409.
[11] V. Yuan, H. Fraunfelder, R. W. Harper et al., Phys. Rev. Lett. **57** (1986) 1680.
[12] J. Alberi, R. Hart, E. Jeenicke et al., Can. J. Phys. **66** (1988) 542.
[13] W.M. Snow, A. Bazhenov, C.S. Blessinger et al., Nucl. Instr. Meth. A **440** (2000) 729.
[14] P.A. Krupchitsky, Fundamental Research with Polarized Slow Neutrons, Springer Verlag Berlin Heidelberg, 1987.
[15] M.M. Nesterov, I.S. Okunev, JETP Letters, **48**(11) (1988) 621.
[16] V.A. Vesna, Yu.M. Gledenov et al., Phys. Atom. Nucl. **62** (1999) 522.
[17] Yu. M. Gledenov, I.S. Okunev et al. Nucl. Instr. and Meth., A **350** (1994) 517.
[18] V. M. Lobashev. Phys. Atom. Nucl. **5** (1965) 957.
[19] V. M. Lobashev, V. A. Nazarenko et. al. Phys. Lett. B **25** (1967) 104.
[20] N.V. Borovikova et al., JETP Letters, **30**(8) (1979) 495.
[21] V.A. Vesna, Yu.M. Gledenov et al., Phys. Atom. Nucl. **59**(1) (1996) 19.
[22] H. Abele, D. Dubbers, H. Häse et al., Nucl. Instr. and Meth., A **562** (2006) 407.
[23] V.A. Vesna, E.A. Kolomensky et al., Nucl. Phys., A **352**(1) (1981) 181.
[24] V.A. Vesna, E.V. Shulgina, PNPI Preprint-2425 (Gatchina 2001), (in Russian).
[25] V. A. Vesna, Yu. M. Gledenov et al., PNPI Preprint-2479 (Gatchina 2002), (in Russian).
[26] V.A. Vesna, Yu.M. Gledenov et al., JETP Letters, **82**(8) (2005) 463.
[27] V.A. Vesna, Yu.M. Gledenov et al., PNPI Preprint-2697 (Gatchina 2006), (in Russian).
[28] Y. G. Abov et al., Nucl. Phys., **34** (1962) 505.
[29] A. Antonov, V.A. Vesna et al., Phys. Atom. Nucl., **48**, 2(8) (1988) 305.